# An updated analysis of satellite quantum-key distribution missions

Olivia Lee[1], Tom Vergoossen[2]

[1]*Raffles Institution (Junior College), 1 Raffles Institution Lane, Singapore 575954*
[2]*Centre for Quantum Technologies, National University of Singapore, Blk S15, 3 Science Drive 2, Singapore 117543*

## Abstract

Quantum key distribution (QKD) is a cryptographic method enabling two parties to establish a private encryption key. The range of communication of ground-based QKD is limited to an order of 100km, due to in-fibre attenuations and atmospheric losses, and the development of quantum repeaters remains technologically challenging. While trusted-node links make communication over large distances possible, satellite-QKD is required for communication over global distances. By using satellites equipped with high-quality optical links, satellite-QKD can achieve ultra-long-distance quantum communication in the 1000-km range. The significant potential of satellite-QKD for the creation of global quantum networks thus makes it a particularly interesting field of research. In this analysis, we begin with an overview of the technical parameters of performing satellite-QKD, including infrastructure and protocols. We continue with a high-level summary of advancements in satellite-QKD by analysing past, present and proposed satellite-QKD missions and initiatives around the world. We conclude by discussing the technical challenges currently faced in satellite-QKD, which can be tackled through future research in this area.

# I. Introduction

Quantum Key Distribution (QKD) is a scheme enabling two parties to derive a private and symmetric encryption key. QKD has the potential to significantly advance information security and encryption processes for mankind. The development of quantum computers represents an increasing threat to conventional public key distribution, driving research into new forms of "quantum-safe" encryption techniques [1-2]. QKD is a promising alternative to public key cryptography, providing unconditional security that cannot be obtained by classical cryptographic means as it is founded on the principles of quantum mechanics [3]. QKD is fundamentally based on the fact that quantum information is coded into the degrees of freedom (e.g. polarization states) of individual photons, which prevents successful attempts to measure and clone a quantum bit (in this case a photon). This is due to the fact that it is impossible to clone a quantum state without irreversibly altering its state (also known as the no-cloning theorem) [4], causing it to lose its information. Hence, attempts by third parties to eavesdrop will necessarily lead to detectable errors, making QKD highly secure and resistant to interference.

The general concept behind satellite-QKD is as follows. A trusted satellite conducts QKD with ground stations to establish independent secret keys with each station. To create a common key to be used by a pair of ground stations (for example stations A and B), the satellite broadcasts the bit-wise parity of each individual key, $K_A \oplus K_B$. Since $K_A$ and $K_B$ are independent strings known only to each station, their bit-wise parity does not reveal any useful information to eavesdroppers. Each station can obtain the other's key: For station A to obtain station B's key, it would determine the bit-wise parity of its own key with the broadcast ($K_A \oplus (K_A \oplus K_B) = K_B$). Similarly, station B can obtain station A's key through the operation $K_B \oplus (K_A \oplus K_B) = K_A$.

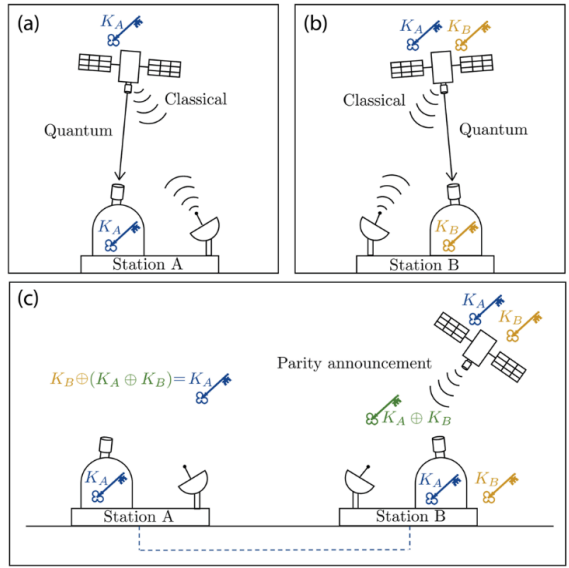

*Fig. 1: Illustration of the most common satellite-QKD scheme: the flying trusted-node. In step (a), the satellite establishes a shared secret key $K_A$ with station A by running a QKD protocol, which requires both classical and quantum communication. This step is repeated in (b) to establish a shared secret key $K_B$, this time with station B located further away. At the end of these steps, the satellite hold both keys, while each station knows only their own. Finally, in step (c), the satellite publicly announces the parity of both keys $K_A \oplus K_B$. This allows station B to determine key $K_A$, which can then be used to encrypt private communication to A and vice versa [101].*

QKD has already been successfully used to distribute secret keys between two parties using fibre networks and free-space ground links [5-7]. Some attempts have even been successful in existing telecommunication infrastructure [8]. However, the main limitation of ground-based QKD is that the maximum communication distance is limited to an order of ~100km [5-7] due to various reasons including atmospheric losses, noise of available single photon detectors, and in-fibre attenuations [9]. Free-space, Earth-bound links are also restricted by the Earth's curvature and atmospheric turbulence [10]. Research has been conducted to develop quantum repeaters (devices that extend the range of quantum communication between sender and receiver) to overcome this problem [11]. However, the development of quantum repeaters remains technologically challenging, as it requires successful entanglement creation over the intermediate distances, as well as storage of the entanglement until entanglement has been established in the adjacent link [12].

There is a need to increase the range of quantum communication for a successful global quantum network to be created. In light of this, a possible solution to this problem is satellite-QKD. Due to low absorption and negligible nonbirefractive character of the atmosphere, optical free space links are superior to fibre links and free-space ground links in achieving ultra-long-distance quantum communication. Hence, optical free space links are currently the most promising channel for large-scale quantum communication by use of satellites and ground stations [13-15]. The usage of a satellite terminal in space makes it possible to develop quantum communication networks on a global scale. Significant experimental efforts have been devoted to investigating the feasibility of satellite-based quantum communications. With the development of new technologies, researchers have extended the communication distance of QKD, entanglement distribution, and quantum teleportation between fixed locations to 1000-km scale [16], which makes satellite-QKD a very promising solution to the technological challenges mentioned earlier [17]. While there are several improvements to be made, satellite-QKD brings us closer to a global quantum network that enables the sharing of strong encryption keys between any two points on Earth.

This report provides an overview of the research and developments that have contributed to the realization of satellite-QKD. We begin by discussing the various technical parameters of performing satellite-based QKD regarding protocols and infrastructure. We continue with a timeline of the key milestones in the development of satellite-QKD, followed by a summary of globally completed and proposed quantum satellite missions, categorized by specific advancements thus far in satellite-QKD. We conclude by discussing the technical challenges faced and future directions to address these challenges.

## II. Technical Parameters of Satellite-QKD

There are several technical parameters that characterise satellite communication links in satellite-QKD, which are expounded below.

*a. Orbit Altitude*

There are three main classes of satellite orbital altitudes: Low Earth Orbit (LEO), Medium Earth Orbit (MEO), and Geostationary Orbit (GEO) (sometimes referred to as High Earth Orbit) [18]. LEO is situated from 180 to 2000 km in altitude, MEO is situated from 2,000 to 35,786 km in altitude, and GEO has an altitude of precisely 35,786 km. For past and current satellite-QKD applications, LEO is the most common option (see Section III.), however future projects might seek altitudes in the MEO or GEO range [19].

Due to the proximity of LEO satellites to the surface, losses due to beam diffraction are significantly reduced. However, there is a tradeoff in the high speed of the satellite relative to the Earth, compromising pointing accuracy during signal transmission and limited time period during which QKD can be performed. The converse is true for satellites of higher altitudes: While the satellite is moving more slowly than in LEO (in the case of MEO) or at rest relative to the ground (in the case of GEO) hence enabling QKD to be conducted continuously, much higher losses are experienced when performing QKD at significantly higher altitudes as free space losses increase quadratically with distance.

Polar LEO will pass ground stations at the poles with every orbit, whereas equatorial LEO do so on the equator [18]. Other orbital inclinations will result in less regular ground station passes, and the time of day of flyovers will vary unless the orbit is sun-synchronous. Satellites in GEO are located above the equator and closer to the horizon approaching the poles, disappearing below the horizon at 81° latitude. Hence, the optical link traverses a much larger amount of atmosphere and will suffer additional losses. To provide polar regions with near-constant satellite coverage, such locations might adopt less conventional orbit choices, such as the Molniya highly elliptical orbits [20].

*b. Link Configuration*

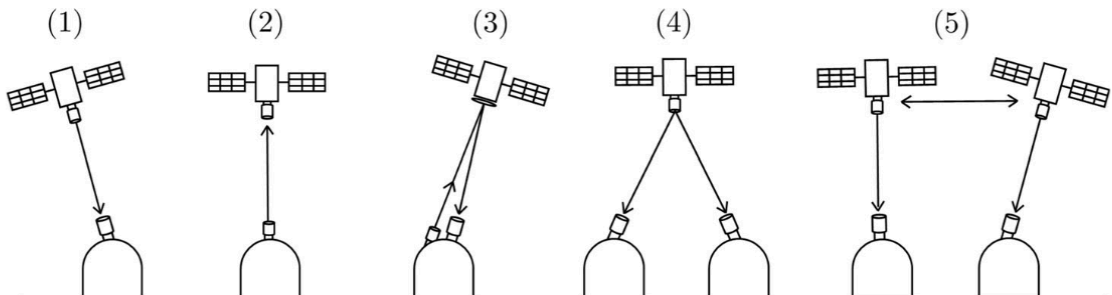

*Fig. 2: Illustration of different platforms for performing satellite-QKD. Scenarios (1) and (2) depict a downlink and an uplink respectively, while in scenario (3) a downlink is simulated by using a retro-reflector on board the satellite. In (4) pairs of entangled photons are being transmitted to Earth so that two ground stations can share entangled states. Finally, scenario (5) illustrates how inter-satellite links can allow more complex satellite-QKD networks [101].*

Quantum communication links with a satellite can be classified either as uplinks or downlinks. This results in several possible configurations for performing QKD with satellites depending on the types of links that are used [21-23].

In a downlink configuration (Scenario 1), the satellite sends signals to the ground. Downlinks are usually recommended for operational QKD, and the only type of link that has been demonstrated [24]. The primary source of optical losses is beam diffraction, which increases with the square of link length, and downlinks always have lower losses for any ground-satellite segment. This arises because atmospheric properties such as turbulence cause the optical beam to wander, which translates into a less accurate ground transmitter compared to a space-based transmitter.

In an uplink configuration (Scenario 2), the ground station sends signals to a receiver in space. The main advantage of using an uplink is that it is not necessary to locate a quantum light source in space, but only to place a receiver on board the satellites [25]. There is a lower photon detection rate on board the satellite and hence a significantly smaller amount of data to be stored and exchanged via a classical, authenticated but non-secure communication channel. It also makes attacks that target receivers significantly more difficult [26]. However, a disadvantage of using uplinks is a greater uplink loss of ~10dB due to atmospheric turbulence, which is most significant in the first 20 km above the Earth's surface. These extra losses are significant when weighed against the advantages of locating quantum light sources on the ground. To overcome this challenge, signal-to-noise ratio filters can be used to discard data at high noise levels, reducing the number of photons required for error correction and privacy amplification [27].

Retro-reflectors on the satellite (Scenario 3) can also be used to create downlinks by modulating signals sent from the ground as they bounce off back to a receiver also on the ground [28]. The challenge here is to develop fast modulating retro-reflectors, and to develop countermeasures that prevent an eavesdropper from sampling the state of the retro-reflector while QKD is being carried out.

In a double downlink configuration (Scenario 4), a source of entangled photon pairs located in the satellite transmits photon pairs to the ground, with each photon in a pair being transmitted to each ground station in communication [29]. This configuration allows the realization of entanglement-based QKD directly between the ground stations, without using the satellite as a trusted node.

Inter-satellite links (Scenario 5) allow for communication between two satellites in the same orbit (intra-orbit links between satellites in LEO, MEO or GEO) or in different orbits (inter-orbit link) [30]. A combination of inter-layer and intra-layer links, called constellations of satellites (see Section IV.) can enable more complex satellite-QKD architectures by allowing each system to collaborate with other systems. This increases the quality of satellite services, decreases the unavailability of services and integrates different services in a single system [31].

*c. DV-QKD or CV-QKD*

QKD protocols can be divided into two categories: discrete-variable quantum key distribution (DV-QKD) or continuous-variable quantum key distribution (CV-QKD). In DV-QKD, information is encoded onto discrete degrees of freedom of optical signals. In CV-QKD, information is encoded in the quadratures of randomly selected coherent states and measured using either homodyne or heterodyne detection [32-33]. While most satellite QKD projects have chosen to implement discrete-variable schemes, significant research has been dedicated to both approaches leading to increasing key generation rates and improved compatibility with current communications infrastructure. However, both approaches are limited in that physical communication channels introduce transmission losses that increase exponentially with distance, greatly limiting the secure key rates that can be achieved over long ranges. There is also a need to reduce the reliance on classical communication between the two parties when establishing the final key [33-34].

*d. Photon Sources*

For DV-QKD, there are two main photon sources: weak coherent pulses (WCP) or polarization-entangled photon-pairs.

Short attenuated pulses from laser diodes provide controlled weak coherent pulses needed to provide photon states for DV-QKD to enhance the security of these systems. However, each pulse has a finite probability of containing more than a single photon [35-36]. Decoy-states have hence been created to reduce the likelihood of photon-number splitting attacks and thus detect eavesdropping. One party randomly chooses between two intensities of coherent state signals, which is revealed publicly to the other party after quantum communication, improving the tolerance to losses compared to the typical prepare-and-measure BB84 protocol that does not employ decoy-states. This increases the transmission distance and rate of key generation [37]. To address the need for active polarization manipulation, the usage of four laser diodes in a single transmitter [35] exploits the high degree of polarization of the diodes, allowing each diode to be identified with a unique polarization state. Using a single laser diode coupled to four waveguides, the side-channel accessible by potential eavesdroppers was closed. Each waveguide was capable of a fixed amount of polarization rotation and the signals were then recombined to result in a single-mode output with four possible polarization states.

Entanglement-based QKD schemes require the generation of photons using polarization entangled photon-pair sources. These sources are based on bulk-crystal, collinear, spontaneous parametric downconversion (SPDC), either periodically-poled potassium titanyl phospate (PPKTP) or single-domain crystals such as beta barium oxide (BBO). SPDC is a non-linear optical process where a photon spontaneously splits into two other photons of lower energies [38]. In particular, Type II spontaneous parametric down-conversion produces photon pairs that emerge on two cones where the vertically polarized photon is on the upper cone and the horizontally polarized photon is on the lower cone [39]. A free-space link is then used to distribute one photon from the entangled pair to Alice, while the other is transmitted to Bob [39].

*d. QKD Protocols*

Several protocols exist to implement QKD between two parties, which can be subdivided into prepare-and-measure (decoy-state BB84) or entanglement-based protocols.

*d(i). Prepare-and-measure protocol: Decoy-state BB84*

Alice encodes each classical bit into the polarization state of an individual photon before transmitting it to Bob. Alice prepares the photon state by randomly choosing between the Z basis (horizontal, |H>, or vertical, |V>) and X basis ($\pm 45º$ i.e. $\frac{1}{\sqrt{2}}$|H>$\pm$|V>), and assigning the states in each basis the values 0 or 1. After choosing the basis, Alice, using a true active random number generator, randomly selects one of the two states and sends it to Bob, who performs a set of measurements on the incoming signals to retrieve the classical data encoded in their states. Bob, using a passive random choice generator (e.g. beamsplitters), randomly selects one of the two bases and performs a measurement, recording the result as classical bits. This repeats for many signals, after which Alice and Bob announce the basis they have chosen for each measurement in the measurement set on a classical communication channel. They keep the measurements that they obtained using the same basis, and discard the measurements that they obtained using different bases. From the preserved signals, a random subset is selected to determine the relative error. In the situation of a sufficiently low error rate (<11%), error-correcting codes and privacy amplification are applied to obtain the final shared secret key [40-43].

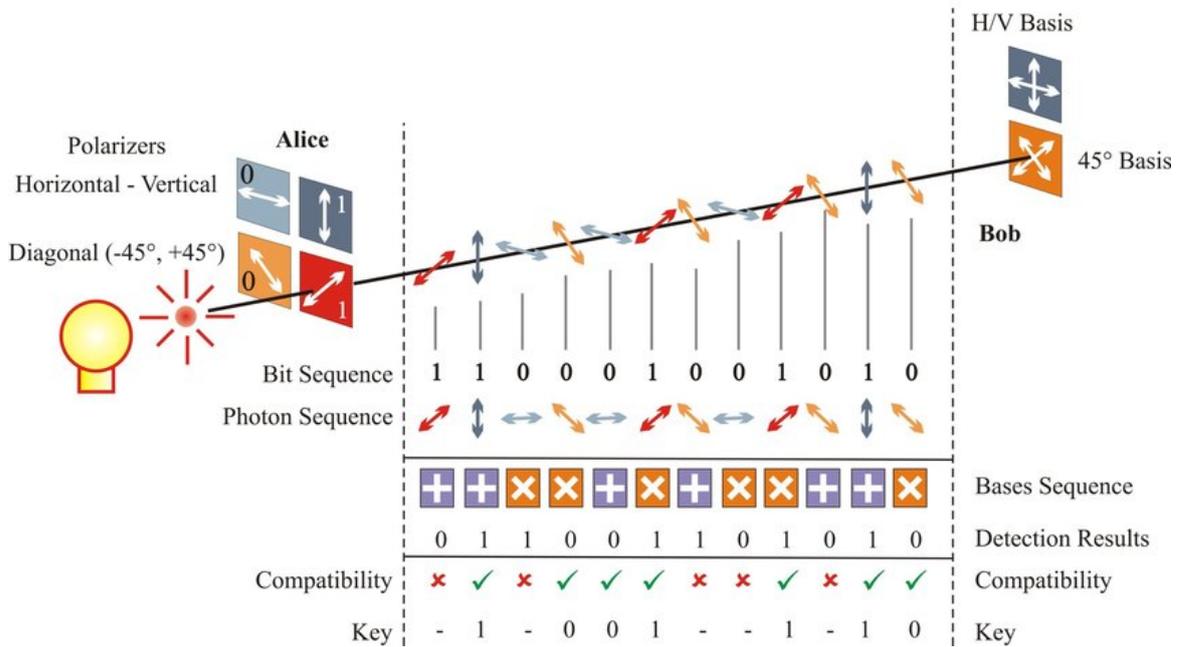

*Fig. 3: Key exchange in BB84 Protocol implemented with polarization of photons [44]*

*d(ii). Entanglement-based protocols*

A maximally entangled photon pair, usually in the polarization degree of freedom, is split such that one photon is transmitted to Alice, while the other is transmitted to Bob. The entangled states are perfectly correlated such that if Alice and Bob both measure their photons with the same basis, they will always get the same answer with 100% probability. Both parties make independent decisions to measure the photons in either the Z or X basis. Since eavesdropping inevitably affects the entanglement between the two photons in an entangled pair, it detectably reduces the degree of violation of Bell's inequality.

In the BBM92 protocol, parameter estimation, error correction, and privacy amplification occur in the same manner as in the BB84 protocol. The main advantage of the BBM92 protocol is that it removes the need for Alice to make an active random choice when encoding states into the photons, and the measurement devices for Alice and Bob are identical [45].

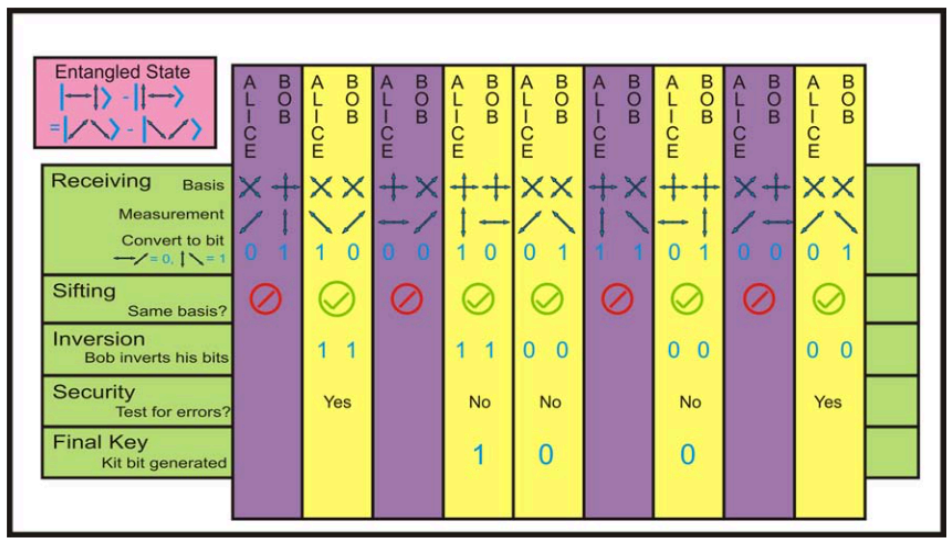

*Fig. 4: Complete BBM92 Protocol. Alice and Bob each receive one photon from a stream of entangled photon pairs, they randomly pick a basis to measure each photon in, get a measurement result, convert their result to a classical bit, sift their results down to only those where they measured in the same basis, use 10% of their measurements to estimate the quantum bit error rate (QBER), and generate a final secure key from the rest of their measurement results [46].*

Alternatively, the E91 protocol [47], through the Bell test, will determine whether the photon-pair correlations between Alice and Bob violate the Bell inequality, confirming the quantum nature of the link and hence its inherent security. After receiving each photon from an entangled pair, Alice and Bob measure the polarization state of every photon in a randomly chosen basis for each photon and note its arrival time. Using a noiseless, authenticated but non-secure public communication channel, Alice and Bob will compare the photon arrival times and the basis in which each photon was measured, preserving the measurements that were conducted using the same basis as the measurement set. By conducting the Bell test on this measurement set, should this correspond to the value expected from Bell's inequality, no local realism was introduced to the system and thus there were no eavesdroppers [48]. This protocol is resistant to eavesdropping as information is only obtained when Alice or Bob perform measurements and key sifting. Eavesdroppers also cannot inject their own data, as doing so would necessarily lead to detection when the Bell's inequality value is too low [49]. After both Alice and Bob obtain the sifted key, similar error correction and privacy amplification processes will be performed to obtain a quantum secured secret key [50].

The E91 measurement scheme is less efficient in its use of photon pairs as the Bell inequality test requires more polarization settings to be monitored. However, entanglement-based protocols have shown to be more tolerant to loss than prepare-and-measure protocols due to the intrinsic timing correlation between photon-pairs generated in the SPDC process [51]. A typical setup is shown below, using active polarization rotators (PR), polarizing beam-splitters (PBS) and avalanche photodiodes (APD) [49].

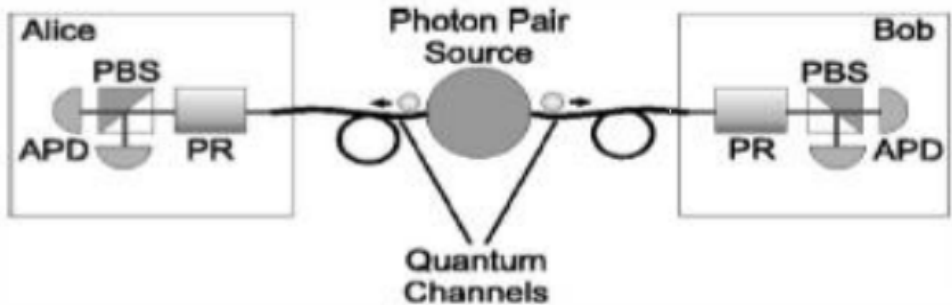

*Fig. 5: System using entangled photon pairs [102]*

*f. Optical Links*

*f(i). Transmitters & Receivers*

Optical links use optical telescopes at the photon source transmitter and at the receiver to beam the photons between satellite and ground station. There are two main types of telescopes: refractive (or transmissive) and reflective [52]. Refractive telescopes use lenses, while reflective telescopes use concave parabolic mirrors. In general, reflective mirrors can be made larger and more durable than lenses. Because only one side of the mirror is used to focus the light, the other side can be placed against a surface for support, enabling a mirror to be much larger relative to a lens. In fact, the largest optical telescope in the world at the Keck Observatory in Hawaii is a reflecting telescope [53]. The larger the collection device, the more light can be directed to the eyepiece, and the brighter the objects appear [54].

However, while the large size of the primary mirror is advantageous for a reflector, this same feature can quickly emphasize the optical aberrations of the telescope. To minimize polarization effects at large angles of incidence, a two-mirror assembly can be used as a tangentially directed one-wave linear retarder. The primary mirror of reflective telescopes acts as a tangentially directed half-wave linear retarder and almost completely depolarizes the linearly polarized component of the light. The secondary mirror, often placed within the path of the beam to block part of the primary mirror, introduces an additional half-wave of linear retardance. Hence, each mirror depolarizes alone but together the two-mirror assembly preserves the polarization state [55].

Comparing space and ground-based telescopes, ground-based telescopes are much larger and consequently can capture more light. Practically speaking, they are less costly, easier to maintain and upgrade, have a much lower risk of being damaged. However, atmospheric distortion is a major issue, and Earth's atmosphere absorbs a lot of the infrared and ultraviolet light that passes through it. While space-based telescopes are smaller, more expensive and difficult to maintain, they can detect frequencies and wavelengths across the entire electromagnetic spectrum [56].

To establish the optical link, a coarse level of mechanical pointing between the satellite and ground station is achieved via satellite orbit determination data (including radar tracking, GPS and star tracker measurements). Two orthogonal axes of rotation are required to track an object across the sky. By mounting the telescope on a two-axis gimbal, one of three gimbal configurations can be used to achieve coarse level pointing for larger spacecraft: altitude over azimuth (Alt-Az), altitude over altitude (Alt-Alt), equatorial, or a combination of the three [57]. For nanosatellites, the entire satellite is usually reoriented. Optical beacons on the ground and satellite can be used for fine-pointing, which is further enhanced by optical beam-steering systems that account for atmospheric turbulence.

*f(ii). Quantum Bit Error Rate (QBER)*

Quantum bit error rate (QBER) is the percentage of the sifted raw key that does not match between Alice and Bob. QBER is generally a direct measure for the secrecy of Alice and Bob's strings since any eavesdropping strategy would perturb the correlations between them [58]. Once QBER exceeds the threshold of 11% QKD protocols based on BB84 will be aborted. The links are where the largest losses from noise and background photons from stray light occur and thus have the biggest impact on the QBER.

Stray light can be minimized through extensive optical blacking measures, however even after extensive filtering [59], it is advised that DV-QKD is conducted at night until better solutions are found [60-61]. It is possible to achieve this by performing DV-QKD at other wavelengths with alternative detectors [62] if suitable light sources can be found. On the other hand, CV-QKD can be conducted during the day as the optical systems used have a sufficiently small spectral bandwidth that allows much of the background to be filtered [63].

*f(iii). Pointing error*

To establish an optical link, there are three stages of pointing, each with increasing accuracy. In the first stage, to achieve a broad level of mechanical pointing between the satellite and the ground, the satellite's orbit must be determined using radar tracking, GPS or star tracker measurements, and this data is exchanged through radio frequency links. In the second stage, a more specific level of mechanical pointing is achieved using laser beacons both on the ground and the satellite. In the third and final stage, the finest level of pointing can be achieved through optical beam-steering systems that serve to correct for atmospheric turbulence.

As mentioned in Section IIb., transmitter pointing accuracy is less significant in uplink than downlink configurations due to atmospheric turbulence. A 2 μrad rms error in the pointing of a 20 cm downlink transmitter would introduce 4 dB of loss compared with <1 dB for a 20cm uplink transmitter. Jitter and imperfections in the tracking systems should be minimized so their contributions to beam broadening are much less than those caused by diffraction and atmospheric turbulence. The receiver system need only point to an accuracy within its field of view, e.g., 50 μrad [64].

*f(iv). Wavelength of signal photons*

There are 2 major wavelength regimes for a quantum channel, depending on the level of expected background photons: 800nm (near the peak quantum efficiency of detectors) and 1500nm (telecom wavelength, traditionally used in optical fibre for its minimum transmission losses). Many optical link systems are limited by the availability of suitable laser sources, specifically in the availability of laser wavelengths. For optical links using SPDC sources, the limiting factor is not photon generation, but the ability of the single photon detectors to distinguish between the photons arriving with small timing separations. Thus to reduce time jitter, the avalanche volume of the photodiode and hence detection efficiency is compromised [65]. Wavelength conversion efficiency is also another constraint as that of SPDC is low; the highest conversion efficiency obtained is $4 \times 10^{-6}$ [66]. Wavelength determination in terms of channel considerations is based on wavelength-dependent losses such as atmospheric absorption, diffraction losses and detection efficiency.

# III. Advances in Satellite-QKD & Summary of completed and proposed missions

Below is a timeline and summary of completed and proposed satellite-QKD missions, organised by general areas of advancement in satellite-QKD.

*Timeline*

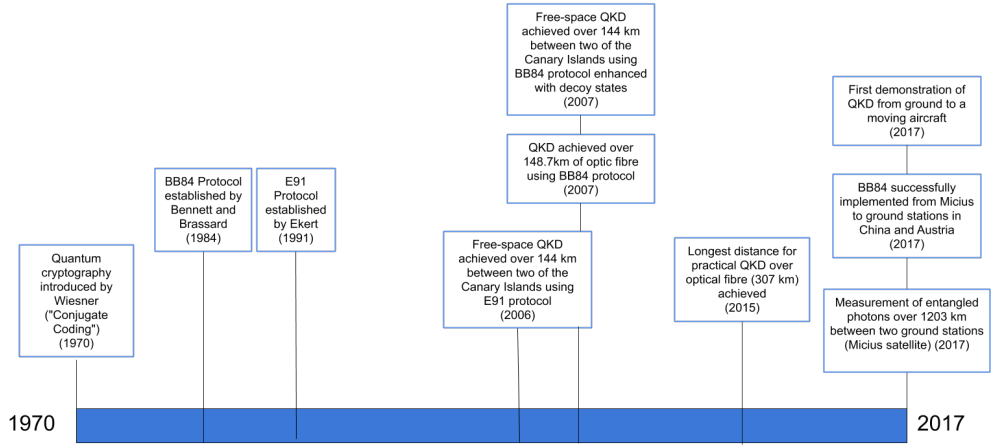

*Advancement 1: Creating a highly accurate optical link (<5 mrad pointing error)*

| Mission | Aim | Satellite | Key Findings | Status, Date |
|---|---|---|---|---|
| Toyoshima et al. [67] | Measuring polarization characteristics of artificial laser source in space through LEO-to-ground atmospheric transmission paths | OICETS 570kg satellite | Polarization preserved within rms error 1.6°. Degree of polarization 99.4±4.4% through space-to-ground atmosphere | Completed, 2009 |
| Yin et al. [68] | Space-ground transmission of quasi-single-photon source using retroreflectors in LEO satellite | CHAMP 500kg satellite | Signal-to-noise ratio of 16:1, sufficient for quantum links for unconditionally secure QKD | Completed, 2013 |
| Vallone et al. [69] | Preservation of polarization state from LEO-to-ground using satellite corner cube retroreflectors as quantum transmitters in orbit | Jason-2 510kg satellite Larets 21kg satellite Starlette 48kg satellite Stella 48kg satellite | Experimental realization of quantum communication from several satellites acting as quantum transmitter and with the MLRO as the receiver. QBER 6.5% | Completed, 2015 |
| QUBE [70] | Testing essential systems for quantum key generation, optical downlink (LEO-to-ground), and attitude control | 3U CubeSat | Once operational in orbit, many new conclusions can be drawn. Follow-up missions will improve QKD in multi-satellite systems and global availability of data encryption through quantum communication. | Proposal, 2018 |

*Advancement 2: Creating sources compatible with launch and space environments*

| Mission | Aim | Satellite | Key Findings | Status, Date |
|---|---|---|---|---|
| QUESS [71] | Establish space platform with long-distance satellite and ground quantum channel | Micius 600kg satellite | Entanglement distribution of 1203 km, teleportation up to 1400 km and BB84 QKD up to 1200 km. QBER ~1%, sifted key rate 14kbps | Completed, 2012 |

*Advancement 3: Versatility - Ability to conduct multiple QKD protocols*

| Mission | Aim | Satellite | Key Findings | Status, Date |
|---|---|---|---|---|
| NanoBob [72] | Photon entanglement uplink experiment. Full quantum communication link >500 km. Versatile payload compatible with a variety of protocols. | 12U CubeSat | Entangled photon source on ground, space segment contains "Bob" detection system only, possible implementation in 12U CubeSat. Receiver compatible with multiple QKD protocols and other quantum physics experiments | Proposal, 2017 |
| QEYSSat [86] | Encryption key generation through quantum uplinks, involving quantum receiver onboard satellite that measures quantum signals sent from photon sources on ground. Investigate long-distance quantum entanglement. | - | - | Funded Mission |
| NanoQEY [103] | Demonstrate long-distance QKD between two distant ground stations on Earth using optical uplink. Distribute ≥10 kbit of secure key between two ground stations, with satellite as trusted node. Perform Bell tests for entangled photons between ground and space | NEMO (Nanosatellite for Earth Monitoring and Observation) (QKD receiver designed by Institute for Quantum Computing, University of Waterloo) | - | Proposal, 2014 |

*Advancement 4: Increasing cost-efficiency by reducing size*

| Mission | Aim | Satellite | Key Findings | Status, Date |
|---|---|---|---|---|
| Takenaka et al. [73] | First downlink microsatellite quantum communication experiment (50kg-class) | SOCRATES 48 kg satellite | Feasible. <5% QBER and 99.4 ± 4.4% degree of polarization with rms error | Completed, 2017 |
| Tang et al. [74] | In-orbit photon counting experiment measuring polarization correlation between photon pairs in nanosatellite | Galassia 1.65kg 2U CubeSat | In-orbit photon correlations exhibit a contrast of 97 ± 2%, matching ground-based tests | Completed, 2016 |

*Advancement 5: Increasing range*

| Mission | Aim | Satellite | Key Findings | Status, Date |
|---|---|---|---|---|
| Günthner et al. [75] | Preservation of quantum coherence properties of quantum states in GEO (propagation over 38,600km to ground) | Alphasat I-XL 6649kg satellite | Quantum limited states arrive at the ground station despite long propagation path. Upper bound for atmospheric and technical excess noise of 0.8 ± 2.4 | Completed, 2017 |
| Dequal et al. [76] | First single photon exchange spanning a distance exceeding 7000 km (MEO) | LAGEOS-2 411kg satellite | Feasible. QBER ~3.6%, signal-to-noise ratio 1.5, 3 count/s. Stable photon transmission for most of passage of MEO satellite. | Completed, 2016 |

*Advancement 6: Moving Receiver Platforms*

| Mission | Aim | Vehicle | Key Findings | Status, Date |
|---|---|---|---|---|
| Nauerth et al. [77] | Ascertain feasibility of BB84 QKD between ground station and airplane moving at angular velocity similar to LEO satellite | Dornier 228 utility aircraft moving at 290km/h | Feasible using advanced pointing tracking system. Compensated for mutual rotations of sender/receiver. QBER 4.8% at 20km range, sifted key rate 145 bit/s, angular speed of 4 mrad/s | Completed, 2013 |

| Bourgoin et al. [78] | First demonstration of optical QKD link to moving receiver platform (truck) traveling at equivalent angular speed of satellite at 600km altitude | Pickup truck (receiver) driven at 33km/h (angular speed of 13 mrad) ~650m from transmitter | Feasible. QBER 6.55%. Sifted key rate 40 bit/s, angular speed. Mean angular deviation after stabilisation of 0.005º at transmitter | Completed, 2015 |
|---|---|---|---|---|
| Wang et al. [79] | Assessing performance under rapid motion, altitude change, vibration, random movement of satellites in high-loss regime using floating hot-air balloon platform | Hot-air balloon with angular velocity 10.5 mrad/s, average angular acceleration at 1.7 mrad/s$^2$ | Feasible. QBER 4.04%, sifted key rate 48bit/s at 96km range. Fine tracking accuracy < 5 μrad | Completed, 2013 |
| Pugh et al. [104] | Investigate utilization of intrinsic strong correlation between pump and output photon spatial modes in SPDC process to mitigate negative effects of atmospheric beam wander. | - | Demonstrates viability of QEYSSAt payload. Prototypes on aircraft and ground close to requirements for satellite uplink. | Completed, 2016 |

*Advancement 7: Polarisation-entangled photon pairs*

| Mission | Aim | Satellite | Key Findings | Status, Date |
|---|---|---|---|---|
| SpooQySats [80] | Establish space worthiness of highly-miniaturised, polarisation-entangled, photon pair sources | 3U CubeSat nanosatellites | Can enable highly secure uplinks and downlinks, symmetric encryption key sharing between optical ground stations | To be launched from ISS in June 2018 |
| CQuCom [81] | LEO-to-ground transmission of entanglement and QKD | 6U CubeSat | High pointing accuracy required to minimize link loss | Proposal, 2017 |

*Advancement 8: Performing free-space daylight QKD*

| Mission | Aim | Satellite | Key Findings | Status, Date |
|---|---|---|---|---|
| Liao et al. [82] | Overcome noise from sunlight, demonstrate free-space daytime QKD over 53 km | - | Feasible, compatible with ground fibre networks.<br><br>Total channel of ~48 dB for 1550 nm free-space working wavelength | Completed, 2017 |
| Gong et al. [83] | Experimentally demonstrate free-space satellite-to-ground QKD in the presence of urban daylight. Develop stochastic parallel gradient descent (SPGD) algorithm with deformable mirror to improve signal-to-noise ratio. | - | Feasible.<br><br>8 km QKD experiment demonstrated over 7 hours throughout daytime via an intra-city free-space link. Final secure key rate of the QKD is 98~419 bps throughout the majority of the daylight hours. | Completed, 2018 |
| Peloso et al. [84] | Demonstrated the continuous running of free-space entanglement QKD system over several full day-night cycles in variable weather conditions | - | Feasible.<br><br>Combination of filtering techniques used to overcome the highly variable illumination and transmission conditions. Continuously generate error corrected, privacy amplified key at an average rate of 385 bps | Completed, 2008 |
| Ko, et al. [85] | Analyze QBER and key rates for different combinations of filtering techniques in a free-space BB84 QKD system operating over 275m in daylight. Optimize conditions of filtering techniques to obtain the maximum secure key rate. | - | QKD system that generates nighttime secure key rate of 341.17 kbps could not generate secure keys in daylight. Noise-filtering combination of 1 nm spectral BPF, 2.5 ns temporal selection, and spatial field of view (FOV) of 566 µrad used. Daytime secure key rate of 191.11 kbps can be achieved using additional filtering techniques with a signal window of 1.75 ns and an FOV of 283 µrad. | Completed, 2018 |

*Advancement 9: Increasing feasibility of uplink transmission*

| Mission | Aim | Satellite | Key Findings | Status, Date |
|---|---|---|---|---|
| QEYSSat [86] | Minimize losses in uplink QKD. Conduct fundamental tests of long-distance quantum entanglement | Microsatellite | Involves quantum receiver onboard satellite that measures quantum signals sent up from the ground | Proposal, 2015 |
| NanoBob [72] | Photon entanglement uplink experiment. Full quantum communication link >500 km. Versatile payload compatible with a variety of protocols. | 12U CubeSat | Tradeoff between uplink losses and versatility. Higher uplink loss (~10 dB) accompanied by lower photon detection rate on satellite and significantly less data to be stored and exchanged | Proposal, 2018 |
| Schiedl et al. [87] | Uplink entanglement-based quantum optics experiments (ground-to-space) | International Space Station | Quantum link can be maintained for 20–70s within one orbital pass. Violation of Bell inequality by 3 standard deviations of statistical significance possible within one satellite pass ($>10^3$ coincidences identified). | Proposal, 2013 |
| $Q^3$Sat [88] | Demonstrate feasibility of establishing Q.Com uplink (ground-to-LEO) with 3U CubeSat. Leverage latest advancements in nanosatellite body-pointing to show that CubeSat can generate quantum-secure key | 3U CubeSat | Feasible. Q.Com achieved over thousands of kilometers via single trusted node. Relatively cheap (<200,000 €). Ignoring finite key effects, pair of ground stations can exchange 13 x $10^6$ secure bits a year. Can be used for Bell tests, clock synchronization, measuring light pollution, earth/atmosphere observations at beacon wavelengths, and studying effect of gravity on quantum systems | Feasibility Study, 2018 |

*Advancement 10: Others*

| Mission | Aim | Satellite | Key Findings | Status, Date |
|---|---|---|---|---|
| CAPSat [89] | Demonstrate active liquid cooling system. Simultaneously reduce vibrations and change pointing direction. Develop single-photon annealing technique to extend the lifetime of radiation-damaged APDs | 3U CubeSat | Investigate 3 experiments utilizing on-board payloads: strain-actuated deployable panels Active thermal system for small spacecraft Single-photon avalanche detectors | Proposal, 2016 |
| SpaceQUEST [90] | Investigate novel concepts for space communication systems. Propose experiments for demonstration of fundamental principles of quantum physics that will benefit from environmental conditions in space (lack of atmospheric disturbance/absorption allowing very long propagation distances) | International Space Station | Quantum communication protocols for space applications: 1. QKD using single and entangled photons 2. Quantum state teleportation 3. Quantum dense coding 4. Quantum communication complexity | Proposal, 2008 |
| Zeitler et al. [91] | Communicating single-photon ququart states from LEO-to-ground between two remote parties through SuperDense Teleportation using hyperentangled photon pairs | International Space Station | Average fidelity of 87.01%, ~2x 44% classical limit. Compared to quantum teleportation and remote state preparation, requires less classical information and fewer experimental resources. Exponentially larger state space volume than the lower dimensional general states with the same number of state parameters. | Proposal, 2015 |
| QuCHAP-IDQuantique [92] | Establish quantum-safe communication networks based on high-altitude platforms | High-altitude platforms | - | Ongoing, 2015 |
| QKDsat [93] by Arqit | Launch a constellation of small low-cost low Earth orbit satellites equipped with QKD payloads | - | - | Proposal, 2018 |
| Qubesat [94] by CQT and RAL | Use QKD technology to test the secure distribution of cryptographic keys between Singapore and the UK | - | - | Funded Mission |

# IV. Challenges and Future Directions of Satellite-QKD

While the above missions have enabled significant advancement of satellite-QKD, there are still a number of technical challenges that can be tackled with further research and future missions.

Firstly, research can be conducted to investigate further reductions in satellite size. With microsatellites (wet mass from 10 to 100 kg) and nanosatellites (wet mass from 1 to 10 kg), further advancements can be made to create a highly accurate optical link with <5 mrad pointing error. There is also a need to create sources compatible with launch and space environments. The above advancements have been achieved with minisatellites (e.g. Micius-type satellites) but not with microsatellites or nanosatellites (e.g. Qubesat). This will not only increase the accuracy and precision of the satellites, but also improve cost-efficiency when the capabilities of microsatellites and nanosatellites are advanced. Reducing satellite size without compromising accuracy hence reduces the cost of a single satellite, and increases the possibility of launching many satellites, enabling the improvement of the spatiotemporal coverage of satellite constellations.

Next, extensions to higher orbits beyond LEO can be investigated. While satellite-QKD missions have been conducted in MEO and GEO ranges (see Section III. Advancement 5), LEO is still the common choice for most applications. LEO is advantageous in its proximity to the surface, thereby reducing beam diffraction losses. Satellites in MEO and GEO are not only at higher altitudes and greater distances from the ground, they must be shrouded to prevent the reflection of sunlight to the ground station. Using brighter transmitters and improving link performance can reduce signal losses, but the performance of MEO and GEO satellites under these conditions will still be less than that of LEO. While LEO is advantageous in its reducing beam diffraction losses, the high speed of satellites in LEO relative to the Earth compromises pointing accuracy during signal transmission and restricts the time period during which QKD can be performed to a small window. Satellites in MEO and GEO, while suffering higher losses, move much more slowly (MEO) or are at rest relative to the ground (GEO), enabling QKD to be conducted continuously. It is necessary to move to higher altitudes as entanglement-QKD implemented using a double-downlink configuration from MEO/GEO can cover very large distances, which greatly facilitates the progression towards building global quantum networks. Concurrent efforts to reduce signal losses in satellites at higher orbits will enable us to reap the benefits of accurate signal transmission and continuous QKD.

The superconducting nanowire single photon detector (SNSPD) is a novel technology that has great potential to advance satellite-QKD missions and quantum technologies involving single-photon detection and advanced photon counting applications. SNSPDs comprise solid-state and optic aspects enabling high-rate (1.3 GBit s-1) quantum key distribution for long-range quantum communication (>1200 km) as well as space communication (239,000 miles). SNSPDs perform excellently in four single-photon relevant characteristics for UV to mid-IR wavelength ranges: high detection efficiency, low false-signal rate, low uncertainty in photon time arrival and fast reset time, though they cannot be optimized simultaneously. They often outperform the best available semiconductor based single-photon detectors [95]. This makes them attractive for several applications in space technologies, nearly commercial quantum systems and quantum optics [96]. However, SNSPDs are currently limited by the availability of cooling technologies, and all past experimental demonstrations have had ground-based applications [97]. This limits their application to satellite-QKD missions using downlink configurations where photon detectors are located at one the ground stations. Further development in SNSPDs allowing them to operate in space can allow them to be incorporated in a greater variety of satellite-QKD missions, specifically those involving uplink configurations.

Future research can also look into the consolidation experimental evidence of different link types facilitating comparative analysis. There have been several missions investigating the feasibility of uplink transmission (see Section III. Advancement 8). Consolidation of experimental evidence of different link types will allow us to directly compare the accuracy of optical links with various configurations.

The creation of constellations of QKD satellites can bring us closer to large-scale or even global quantum networks that enable the sharing of symmetric encryption keys between any two points on Earth. A constellation model is described which enables QKD-derived encryption keys to be established between any two-ground stations with low latency [98]. To provide global real-time quantum communication connectivity, a feasible solution is the building of a satellite constellation (SC), composed of multiple quantum satellites operating in LEO, and high-earth-orbits (HEO), including GEO satellites [99]. Investigations into regional and eventual global networks can be made, as well as considering the use of inter-satellite QKD links for the transfer of keys between LEO trusted-note QKD satellites, GEO relay satellites and ground stations.

Additionally, comparing DV-QKD and CV-QKD protocols is a potential area for further research. Historically, most missions, including those discussed in Section III, make use of discrete-variable schemes such as photon polarization, where information is encoded onto discrete degrees of freedom of optical signals. Further investigations can be made into CV-QKD, which involves encoding information in the quadratures of randomly selected coherent states (such as position or momentum). Comparisons of the performance of the two classes of protocols can be made using metrics such as relative secret key rates, transmission distances, communication overheads and computing resource requirements of error correction codes.

Finally, while daytime QKD has been experimentally proven to be feasible from space, further investigations into daytime QKD from the ground can be made. A significant challenge of implementing free-space QKD systems in daylight is the impact of scattered background noise photons from sunlight. This hence requires elaborate elimination in spectral, temporal, and spatial domains to decrease the QBER and guarantee the system's security [100]. While space-based daylight QKD has been relatively successful, different combinations of filtering techniques can be investigated to increase the accuracy of ground-based daytime QKD.

In conclusion, significant progress has been made in satellite-QKD in recent years. While there are still a number of technological challenges to overcome through further research, many advances have been made thus far, bringing us closer to robust, accurate and quantum-safe encryption methods, and potentially a global quantum network in the near future.